# Cluster glass behavior and magnetocaloric effect in the hexagonal polymorph of disordered $Ce_2PdGe_3$

Leszek S. Litzbarski[*,1], Kamil Balcarek[1], Anna Bajorek[2], Tomasz Klimczuk[1], Michał J. Winiarski[1], Karol Synoradzki[3]

[1]Gdansk University of Technology, Narutowicza 11/12, 80-233 Gdańsk, Poland

[2]A. Chełkowski Institute of Physics, University of Silesia in Katowice, 75 PułkuPiechoty 1, Chorzów, 41-500, Poland

[3]Institute of Molecular Physics, Polish Academy of Sciences, Smoluchowskiego 17, 60-173 Poznań, Poland

* E-mail: Leszek.litzbarski@pg.edu.pl

In this work, we study the hexagonal variant of the $Ce_2PdGe_3$ system that crystallizes in the $AlB_2$-type structure (space group $P6/mmm$, h$P3$) and exhibits cluster spin glass type behavior. The physical properties were studied by magnetization, heat capacity and electric resistivity, which showed that $AlB_2$-type $Ce_2PdGe_3$ ($h$-$Ce_2PdGe_3$) can be classified as a cluster glass material with the freezing temperature $T_f$ = 3.44 K in contrast to the behavior of the previously described tetragonal variant, which shows a double antiferromagnetic transition at $T_{N1}$ = 11 K and $T_{N2}$ = 2.3 K. The X-ray photoelectron spectroscopy measurements reveal that the Ce 4f states are well localized. In addition, we examine the magnetocaloric effect in this compound. The maximum values of magnetocaloric parameters appear in the vicinity of 7-9 K. For a magnetic field change of 50 kOe, the value of the change in magnetic entropy is 2.6(1) J kg$^{-1}$ K$^{-1}$ and the adiabatic temperature change is ~8 K.

## 1. Introduction

Ce-based compounds exhibit unique electronic, transport, catalytic and magnetic properties due to the presence of cerium, a versatile rare earth (RE) element. Understanding the fundamental aspects of their behavior can provide valuable insights into the broader field of intermetallic compounds and pave the way for the discovery of novel materials with enhanced properties. Depending on the chemical composition and crystallographic structure, Ce-based compounds can exhibit various effects, e.g. superconductivity, the Kondo effect, quantum criticality, various types of magnetic ordering (mainly antiferromagnetic) [1], [2], [3], [4]. Within the myriad of possibilities, our focus was on the $Ce_2PdGe_3$ compound, known for its ability to crystallize in two distinct structures: the hexagonal $AlB_2$-type and the tetragonal α-$ThSi_2$-type.

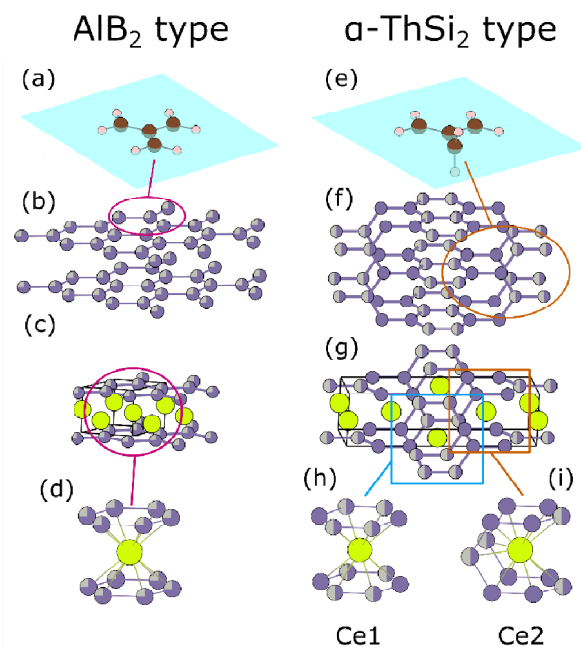

**Figure 1.** (a,e) Two conformations of a trimethylenemethane molecule and the two derived 3-connected network: AlB$_2$ type (b) and α-ThSi$_2$ type (f). Panels (c) and (g) show the crystal structures of *h*-Ce$_2$PdGe$_3$ and *t*-Ce$_2$PdGe$_3$, respectively. The AlB$_2$-type *h*-Ce$_2$PdGe$_3$ features only one inequivalent Ce site with hexagonal prismatic coordination (d). Pd and Ge are distributed uniformly over all hexagonal ring positions. In the *t*-Ce$_2$PdGe$_3$ variant there are two Ce 12-coordinate sites: one with hexagonal prismatic coordination (Ce1; panel (h)) and the other with a capped square prismatic coordination (Ce2; panel(i)).

The hexagonal AlB$_2$-type is one of the simplest and most ubiquitous crystal structures [5]. It consists of graphite-like covalently-bound honeycomb layers separated by triangular layers formed usually by an electropositive element. Numerous related structure types can be derived from the AlB$_2$ aristotype[6].

One of the interesting derivatives of the layered structure of AlB$_2$ is the tetragonal α-ThSi$_2$-type, which shares the same local connectivity (three-connected network) but has a 3-dimensional character [7]. The covalent network of AlB$_2$ and α-ThSi$_2$ can be conceptually derived from the two possible conformations of the trimethylenemethane molecule (Fig. 1 (a,e)), as discussed by Zheng and Hoffmann [7]. Relative stability of the two structural variants depends on electron count (governing the filling of the π$^*$ antibonding orbitals of the covalent network) [7], [8], [9], but also e.g. on atomic packing factors [10]. In the case of the parent compound ThSi$_2$, the two polymorphs can be stabilized at different temperature ranges[11], [12]: the hexagonal AlB$_2$-type β variant of ThSi$_2$ is stable at 850°C, while annealing at higher temperatures yields the tetragonal α-ThSi$_2$[12]. This effect shows that in some cases the relative stability of the variants is comparable.

A vast number of ternary AlB$_2$- and α-ThSi$_2$-type derivatives have been reported [8], [9], [13], [14], [15], [16] among *RE$_2$TMX$_3$* intermetallic compounds (*RE* - rare earth and actinide metals; *TM* – late transition metals; *X* – Si, Ge). In some cases different polymorphs can be obtained by a small alteration of stoichiometry [10], [17], [18], [19], heat treatment [20], or the use of different synthesis techniques [21], [22], [23]. It should be mentioned that the structural transition usually results in significant changes in the physical properties of the *RE$_2$TMX$_3$* compounds.

Strydom reported that $Ce_2PdGe_3$ crystallizes in the hexagonal $AlB_2$-type structure (Fig. 1(b,c)) [21], while an earlier report by Kitagawa *et al.* suggests a $ThSi_2$-type derivative structure (Fig. 1 (f,g)) for compounds with a Ce:Pd:Ge atomic ratio close to 2:1:3 [24]. Detailed studies of the physical properties on single crystals of the tetragonal $Ce_2PdGe_3$ (*t*-$Ce_2PdGe_3$) have been reported by Baumbach *et al.*[23] and Bhattacharyya *et al.*[25], revealing a complex antiferromagnetic behavior with two transition temperatures $T_{N1} \approx$ 11 K and $T_{N2} \approx$ 2.3 K, resulting from the ordering of two Ce inequivalent sublattices. In the tetragonal, α-$ThSi_2$-derived structure of $Ce_2PdGe_3$ (referred to as *t*-$Ce_2PdGe_3$ throughout the rest of the paper), one Ce site (Ce1) is coordinated by two hexagonal $(Pd,Ge)_6$ rings (see Fig. 1(h)), while the other (Ce2) has the same coordination number (12) but sits in the center of a square $Ge_4$ prism capped by four Ge/Pd (mixed) atoms (see Fig. 1(i)) [23]. Meanwhile, the hexagonal variant has only one 12-coordinated Ce site with Pd and Ge randomly distributed over the hexagonal rings (Fig. 1(d)). To date the magnetic properties of the hexagonal $Ce_2PdGe_3$ (referred to as *h*-$Ce_2PdGe_3$ throughout the rest of the paper) have not been reported.

Here we describe the synthesis and characterization of the *h*-$Ce_2PdGe_3$. We report a cluster glass-like magnetic behavior, resulting from the disordered graphite-like network (the triangular Ce lattice remains ordered). We also discuss the differences in magnetic properties of the tetragonal and hexagonal variants of $Ce_2PdGe_3$.

## 2. Materials and Methods

The polycrystalline sample of *h*-$Ce_2PdGe_3$ was prepared using stoichiometric amounts of high- purity constituent elements i.e. cerium (99.9% Onyxmet), palladium (99.95% Alfa Aesar) and germanium (99.999% Alfa Aesar). The synthesis was carried out in the MAM-1 arc furnace by GmbH Edmund Buhler in an argon atmosphere with a Zr-getter. The melting process was repeated several times, each time turning the ingot in order to improve a homogeneity of the synthetized material. The weight losses during a melting were negligible (< 0.5%), which indicates that the relative concentration of chemical elements in the ingot was close to the nominal composition. The resulting samples had a dark silvery color. Selected samples were heat treated at temperatures ranging from 800 °C to 1200 °C for 15, 30, 45 and 60 hours. In order to avoid their oxidation during this process, samples were sealed in quartz tubes filled with a high purity Ar.

The crystal structure of the obtained samples was characterized by powder X-ray diffraction (pXRD) after manually grinding in a mortar to a fine powder. The pXRD patterns were collected at room temperature by a Bruker D2Phaser diffractometer equipped with a XE-T detector (Cu $K_α$ radiation). Diffraction patterns were analyzed using the Rietveld refinement method FULLPROF software [26]. The cast sample turned out to be phase pure (see Fig. 2). We also found that the annealing process results in a formation of impurity peaks and therefore as-cast samples were taken for measurements of physical properties. The Archimedes method was employed to measure density at room temperature, utilizing isopropyl alcohol [$(CH_3)_2CHOH$] and an electronic balance from Radwag (Radom, Poland) model XA 110.4Y.A.

The electronic structure analysis was determined by X-ray photoelectron spectroscopy (XPS) using a Physical Electronics (PHI 5700/660) spectrometer working under an ultra-high vacuum ($10^{-9}$ Torr) in UHV cluster and a monochromatic Al Kα X-ray source (1486.6 eV). The tested sample was initially kept pre-chamber and held under vacuum ($10^{-8}$ Torr) for at least 12 h, then transferred to the preparation chamber and cleaved under UHV conditions of $10^{-9}$ Torr to obtain a fresh surface. Afterwards, the cleaved sample was transferred immediately to the measurements chamber and analyzed. The survey spectra were acquired with a pass energy of 187.85 eV, and the core level lines were measured with a pass energy of 23.5 eV respectively. The obtained spectra were processed with XPSPEAK41 software. The deconvolution of the core level lines was performed by applying the Shirley background type and the Doniach – Sujnić line shape.

The physical properties characterization i.e. magnetization, heat capacity and resistivity measurements, was investigated using a Quantum Design Physical Properties Measurement System (PPMS) equipped with an AC Measurement System (ACMS) and a vibrating sample magnetometer (VSM). Measurements were made on a small piece (~20 mg) of polycrystalline sample with an irregular sphere-like shape. Magnetization measurements were performed for various magnetic field (up to 90 kOe) after field cooling (FC) as well as zero field cooling (ZFC) both in AC and DC mode. The specific heat measurements were recorded with and without an applied magnetic field using a standard thermal relaxation technique in the temperature range $T$ = 1.9 - 300 K. Electrical resistivity measurements were carried out using a four probe technique with platinum wire contacts, which were spot-welded to the polished sample surface. The sample for transport measurements was cut into cuboid shapes using a wire saw.

## 3. Results and discussion

Fig. 2 presents the results of a room temperature pXRD measurements for an as-cast $h$-Ce$_2$PdGe$_3$ sample. Rietveld's analysis indicates that this compound crystallizes in a hexagonal structure with the space group $P6/mmm$ (no. 191) and unit cell parameters ($a$ = 4.2608(2) Å, $c$ = 4.2472(1) Å), which are close to the values reported by Strydom [21]. Similar to other $RE_2TM$Ge$_3$ (e.g. Nd$_2$PdGe$_3$[27], Gd$_2$PdGe$_3$[28] and Tb$_2$PdGe$_3$[18]) compounds there are no superstructure reflections visible in the pXRD pattern, suggesting a statistical disorder within the Pd-Ge plane. The comparable lengths of lattice constants cause that the strength of interactions between the nearest neighbors and the next nearest neighbors are similar, which plays a role in the magnetic frustration [29]. Disorder within the graphene-like layers and geometrical frustration (Ce atoms arrange themselves in a triangular lattice, as in typical frustrated magnets such as YbMgGaO$_4$[30] and delafossites[31]) are essential for the formation of a glassy magnetic state [32].

The measured density value for our sample was 7.31(1) g cm$^{-3}$ and is close to the value determined from the XRD analysis (7.512 g cm$^{-3}$).

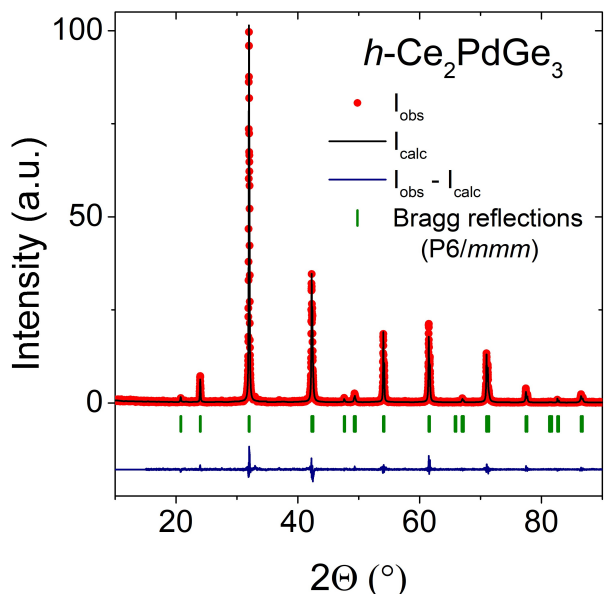

**Figure 2.** pXRD pattern of as-cast $h$-Ce$_2$PdGe$_3$ (red points) with a Rietveld fit (black line) using a AlB$_2$-type structure model. Blue line shows the difference between the measured and calculated intensity, green tick – expected positions of Bragg reflections for the AlB$_2$-type structure. Cell and structural parameters resulting from the Rietveld refinement are gathered in Table 1. Conventional Rietveld $R$-factors (for points with Bragg contribution): $R_p$ = 14.2%, $R_{wp}$ = 19.6%, $R_{exp}$ = 15.85%, $\chi^2$ = 1.53.

XPS is a powerful technique that provides valuable insights into the material under study by identifying its elemental composition and chemical states. It also reveals details about the material's electron structure and electron density distribution. In this study, XPS measurements were performed on the Ce$_2$PdGe$_3$ sample, with the results shown in the supporting information (SI).

The XPS spectra show peaks from only the elements present in the Ce$_2$PdGe$_3$ compound, with minor peaks from carbon and oxygen (Fig. S1). The spectrum near the Fermi level ($E_F$) is dominated by peaks corresponding to Ce 4$f$, Pd 4$d$, and Ge 3$p$ states. The shape of the XPS spectrum near the $E_F$ indicates that the material is a metal.

The XPS spectra for the Pd 3d and Ge 2p states are shown in Fig. S3. Both exhibit a pair of sharp peaks caused by spin-orbit (SO) interaction. For Pd, the Pd 3d$_{3/2}$ and Pd 3d$_{5/2}$ peaks occur at binding energies of 340.8 eV and 335.7 eV, respectively, with an SO interaction energy of 5.1 eV. Ge 2p peaks are located at 1247.4 eV and 1216.3 eV, corresponding to an SO interaction energy of 31.1 eV. These peak positions match well with the literature data [33], [34] and the absence of additional peaks or humps indicates that neither element is oxidized. Additionally, a broad peak around 346 eV, likely due to plasmonic excitations of Pd, is observed. In the Ge 2p spectrum, a C Auger peak and a broad feature associated with Ge losses are also present.

**Table 1.** Comparison of structural parameters for $Ce_2PdGe_3$ crystallizing in a hexagonal and tetragonal structure. $B_{iso}$ is the isotropic thermal displacement parameter $B_{iso} = 8\pi^2<u^2>$, where $<u^2>$ is a mean square isotropic displacement of the atom. SOF is the normalized occupancy factor. Numbers in parentheses are uncertainties of least significant digits.

| $Ce_2PdGe_3$ | Hexagonal (this work) | Tetragonal (ref. [20]) |
|---|---|---|
| Space group | $P6/mmm$ (no. 191) | $P4_2/mmc$ (no. 131) |
| $a$ (Å) | 4.2608(2) | 4.24440(8) |
| $c$ (Å) | 4.2472(1) | 14.7928(2) |
| $V$ (Å$^3$) | 66.77(1) | 266.491(5) |
| $Z$ | 1 | 4 |
| Molar weight (g mol$^{-3}$) | 604.49 | 607.22 |
| Density (g cm$^{-3}$) | 7.512 | 7.567 |
| Atomic coordinates | | |
| Ce | $x = y = z = 0$<br>SOF = 1<br>$B_{iso}$ = 0.83(8) Å$^2$ | Ce1 $x = ½\ y = 0\ z = ½$<br>SOF = 1<br><br>Ce2 $x = ½\ y = ½\ z = ¼$<br>SOF = 1 |
| Pd | $x = 1/3\ y = 2/3\ z = ½$<br>SOF = 0.25<br>$B_{iso}$ = 1.30(1) Å$^2$ | $x = 0\ y = ½$<br>$z = 0.41877(1)$<br>SOF = 0.542(8) |
| Ge | $x = 2/3\ y = 1/3\ z = ½$<br>SOF = 0.75<br>$B_{iso}$ = 1.30(1) Å$^2$ | $x = 0\ y = 0$<br>$z = 0.33323(4)$<br>SOF = 1<br><br>$x = 0\ y = ½$<br>$z = 0.41877(1)$<br>SOF = 0.542(8) |

The XPS spectra for Ce 3$d$ and 4$d$ states are presented in Fig. S4. In the case of the spectrum for the 4d states, two peaks were observed at energies 108 and 110. These two peaks arose from S-O splitting and are assigned to the 4$d^9$4$f^1$ and 4$d^9$4$f^2$ final states. If the 4$f^0$ states are present, two additional peaks should appear in this spectrum at a distance of about 11 eV. Unfortunately, peaks from the Ge 3p states appear in this binding energy range. Thus, from this spectrum, it is not possible to determine the state of the Ce ions in this compound. Fortunately, we can get more information from analyzing the Ce 3d spectrum, which is dominated by two broad maxima that are associated with the 3d$^9$f$^1$ and 4d$^9$4f$^2$ states, which have been split by the spin-orbit interaction ($\Delta_{SO}$ = 18.6 eV). In addition to these peaks located at 880.5 eV, 883.9 eV, 899.1 eV, and 902.5 eV, we observed two broad maxima at ~896 eV and ~914 eV, which were assigned to plasmon losses with an energy of $h\nu \approx$ 13 eV. The detailed analysis of the Ce 3d spectra has been conducted in accordance with the theoretical framework established by Gunnarsson and Schönhammer[35]. Within this framework, the intensity of the $f^2$ peak serves as an indicator of the strength of hybridization; thus, the energy coupling parameter $\Delta$ between the $f$ electrons and conduction bands can be extracted from the intensity ratio $r_2 = I(f^2)/[I(f^1)+I(f^2)]$. The occupancy of the 4$f$ states, denoted as $n_f$, can be approximated using the ratio $r_0 = I(f^0)/[I(f^0) + I(f^1) + I(f^2)]$, as the $f^0$ satellite offers insights into the count of $f$ electrons. A value of $n_f \sim 1$ is indicative of the stable $Ce^{3+}$ state;

conversely, in instances where the Ce ions exhibit fluctuating valence characteristics, $n_f$ is found to be less than 1. The result of fitting is shown in Fig. S4(b). The values of the obtained parameters are Δ = 50(5) meV and $n_f$ ~ 1 and are typical for Ce-based intermetallic compounds with stable valence 3+ [36], [37], [38], [39], [40], [41].In conclusion, the XPS measurements confirm the high quality of the $Ce_2PdGe_3$ sample, with no evidence of oxide formation and Ce ions predominantly in the 3+ oxidation state.

The DC magnetic susceptibility ($\chi \approx M/H$) measurements on $h$-$Ce_2PdGe_3$ were carried out in an applied magnetic field value of $H$ = 1 kOe. The value of $\chi(T)$ gradually increases with decreasing temperature in a manner consistent with the Curie-Weiss paramagnetic behavior. The low temperature region exhibits a cusp which indicates a magnetic transition. In Fig. 4 is presented the temperature dependence of inverse susceptibility $1/\chi$, which is fitted with the modified Curie Weiss law:

$$\frac{1}{\chi} = \frac{T - \theta_{CW}}{\chi_0 (T - \theta_{CW}) + C}, \tag{1}$$

where $C$ is the Curie constant, $\chi_0$ is the temperature-independent susceptibility and $\theta_{CW}$ is the Curie-Weiss temperature. The red line represents the fit of this formula to the linear region of a $1/\chi$ plot ($T$ = 75 - 300 K) which yields a value of $\theta_{CW}$= -3.7(1) K. This result is about two times lower than found for $t$-$Ce_2PdGe_3$ ($\theta_{CW}$= -6.3 K) [23]. The negative value of the Curie temperature suggests an average antiferromagnetic coupling between the magnetic moments.

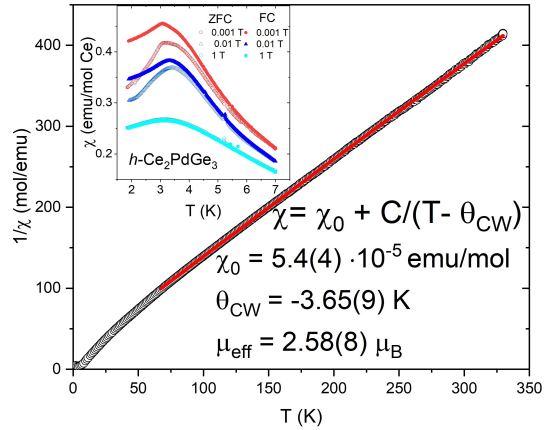

**Figure 3.** The inverse magnetic susceptibility vs. temperature with fitted modified Curie-Weiss law (red line) for $h$-Ce2PdGe$_3$. The inset shows the ZFC and FC curves measured at different external magnetic fields

The Curie constant was employed to estimate the effective magnetic moment using the equation:

$$\mu_{eff} = \left(\frac{3Ck_B}{\mu_B^2 N_A}\right)^{1/2} \tag{2}$$

where $k_B$ is the Boltzmann constant, $\mu_B$ is the Bohr magneton and $N_A$ is the Avogadro number. The experimentally calculated value of $\mu_{eff}$ is almost as same as the theoretically expected free ion moment of 2.58 $\mu_B$ per $Ce^{3+}$ [42],which indicates the trivalent nature of the cerium ion in the studied compound. A similar observation was made for $t$-$Ce_2PdGe_3$[23]. In order to elucidate the nature of magnetic order in $h$-$Ce_2PdGe_3$ the low-temperature $\chi(T)$ measurements in zero field cooling (ZFC) as well as field cooling (FC) conditions for different applied magnetic fields were carried out. Both ZFC and FC curves, which are presented in the inset of Fig. 4, have a well-defined maxima below $T$ = 3.5 K. The observed bifurcation of $\chi(T)$ curves is smeared and becomes negligible with an increasing magnetic field, which suggests a spin glass like transition rather than a long range antiferromagnetic ordering [23]. The value of an irreversibility temperature ($T_{irr}$) was defined as a maximum of $\chi(T)$ for $H$ = 10 Oe and is equal $T_{irr}$ = 3.1 K, which is different from both transition temperatures seen in $t$-$Ce_2PdGe_3$ [23]. This parameter can be used to estimate the empirical measure of frustration employing a criterion proposed by Ramirez ($\phi = |\theta_{CW}|/T_{irr}$) [29]. The obtained value of $\phi$ = 1.2 indicates an occurrence of weak magnetic frustration [29], [43], leading (together with Pd-Ge site disorder) to a glassy system formation.

Clear hysteresis loops were observed in the $M(H)$ relationship for $h$-$Ce_2PdGe_3$ at temperatures below 10 K (Fig. 5). The value of the coercivity field increases as the temperature decreases. Hysteresis is usually observed in ferromagnetic materials but is also typical of spin glasses [44].This phenomenon likely originates from field-induced spinreorientation effects, which can influence the magnetic properties of the system even above the freezingtemperature[45].The measured $M(H)$ curve exhibits nonlinear behavior, two distinct inflections are seen in the magnetic field of ~1 and ~13 kOe (for the primary curve measured at $T$ = 2 K). The maximum recorded value of magnetization determined in $\mu_B$ for the Ce ion at $T$ = 2 K and in a magnetic field of 90 kOe is lower than the theoretical value for the free $Ce^{3+}$ ion ($g_J J$ = 2.14). The reduction may be due to the fact that the value of the magnetic field is too weak to completely reorient the Ce magnetic moments. The reduction in the value of the magnetic moment is often observed in spin-glass materials.

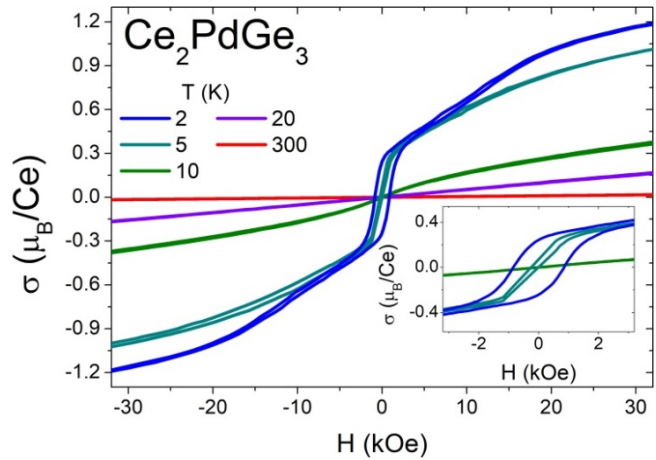

**Figure 5.** Isothermal magnetization as a function of applied magnetic field at different temperatures. Inset shows zoom-in in low magnetic field values.

Further investigations of the magnetism in the $h$-$Ce_2PdGe_3$ were carried out via AC susceptibility measurements with excitation frequencies $\nu$ = 37, 113, 347, 1065, 3263 and 10,000 Hz (logarithmic spacing) and excitation field $H_{ac}$ = 5 Oe. The real ($M'$) and the imaginary ($M''$) part of the ac magnetization is presented in Fig. 7, exhibiting a peak around $T_{irr}$, which is sensitive to an applied frequency and shifts toward higher temperature with increasing frequencies. This phenomenon is a characteristic feature of glassy magnetic materials (such behavior is not observed in antiferromagnets) and can be used to classify $h$-$Ce_2PdGe_3$ as canonical spin glass or cluster glass [46]. Firstly, the observed cusp may be used to estimate a freezing temperature ($T_f$), which is commonly defined as a maximum of the $M$ vs. $T$ curve measured at the lowest frequency [47]. For $h$-$Ce_2PdGe_3$ $T_f$(37 Hz) = 3.44 K. This value is then taken to obtain the relative shift in $T_f$ per decade of the frequency using the relation [44]:

$$\delta T_f = \frac{\Delta T_f}{T_f(37\,Hz)\Delta log \upsilon}. \qquad (3)$$

In the case of $h$-$Ce_2PdGe_3$ $\delta T_f$ = 0.012, which is higher than typically observed in canonical spin-glasses (~$10^{-3}$), however it falls within the range expected for cluster glass compounds e.g. $Nd_2Ni_{0.94}Si_{2.94}$ ($\delta T_f$ = 0.029)[48], $CeCu_4Mn_{0.9}Al_{0.1}$ ($\delta T_f$ = 0.0177) [37]and $Tb_2Pt_{1.2}Ge_{2.8}$ ($\delta T_f$ = 0.046) [19].

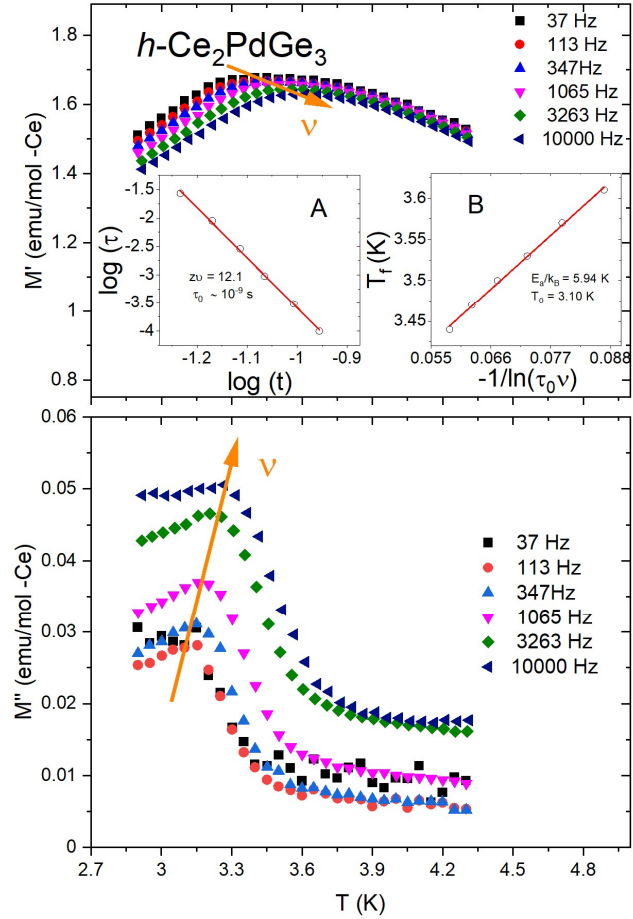

**Figure 6.** Temperature dependence of the real part of the ac magnetic susceptibility $\chi(T)$ for $h$-$Ce_2PdGe_3$. The inset (A) shows ln($\nu$) plotted as a function of ln($t_{SG}$) with the solid red line, which represents the fit to the power-law divergence. The inset (B) shows a plot of the freezing temperature ($T_f$) versus $1/\ln(\tau_0\nu)$ with a Vogel-Fulcher law fit (red solid line).

The second criterion to distinguish the glassy systems is based on a dynamical scaling theory of a critical slowing down and may be described by a power law dependence[46]:

$$\tau = \tau_0 \left(\frac{T_f - T_{SG}}{T_{SG}}\right)^{-zv'}, \qquad (4)$$

where $\tau \sim 1/\nu$ is the relaxation time associated with a measured frequency $\nu$ and $zv'$ is a dynamic critical exponent, which correspond to the correlation length ($\xi = (T_f/T_{SG} - 1)^{-v'}$, $\tau \sim \xi^z$) [49]. $\tau_0$ is a microscopic single spin flipping time (a characteristic flipping time of the magnetic entities) and lie in the range $\tau_0 = 10^{-7}$ s for cluster glass compounds to $\tau_0 = 10^{-13}$ s for spin-glass materials [46]. The last parameter is a spin-glass temperature - $T_{SG}$, which stands for a temperature in the static limit ($\nu \to 0$) and is usually equated with $T_{irr}$. The Inset (A) Fig. 6 displays log($\tau$) - log($t_{SG}$) dependence, where $t_{SG}$ is a reduced temperature defined as $t_{SG} = (T_f(37Hz) - T_{SG})/T_{SG}$. The red line represents a linear fit of the power law to the experimental data, which was used to estimate values of $zv'$ and $\tau_0$. The obtained results ($zv'$ = 12.1 and $\tau_0 \sim 10^{-9}$ s) indicate that $h$-Ce$_2$PdGe$_3$ may be classified as a cluster glass system and are necessary in a further analysis of $\chi'(T)$ measurements. $T_f$ is considered as the activation temperature of thermal interactions between magnetic spins, which dynamics can be modelled empirically using the Vogel-Fulcher (VF) method expressed by equation [46]:

$$\tau = \tau_0 exp\left(\frac{E_a}{k_B(T_f - T_0)}\right), \qquad (5)$$

where $E_a$ is the activation energy and $T_0$ is a measure of the interaction strength between the magnetic spins, known as the VF temperature. This formula may be also rewritten as:

$$T_f = T_0 - \frac{E_a}{k_B} \frac{1}{\ln(\tau_0 \nu)}, \qquad (6)$$

which is exhibited as a red line fitted to the plot in right inset of Fig. 6(a). The values of these parameters determined from the linear fit of $T_f$ vs $1/\ln(\nu_0/\nu)$ graph are equal $E_a/k_B$ = 5.94 K and $T_0$ = 3.10 K, which is an another evidence for a cluster glass formation. In the case of a cluster glass materials the interactions between magnetic entities are rather weak, which is represented by $E_a/k_B T_0$ ratio greater than 1, whilst for canonical spin glasses this ratio should be close to one[47]. Ultimately, our hypothesis agrees with the Tholence criterion defined as [50]: $\delta T_{Th} = (T_f(37\ Hz) - T_0)/T_f(37\ Hz) = 0.10(2)$, which is a value similar to data reported earlier for other $RE_2TM$Ge$_3$ cluster glass compounds e.g. Ho$_2$Pd$_{1.3}$Ge$_{2.7}$ ($\delta T_{Th}$ = 0.32) [10], Dy$_2$Pt$_{1.15}$Ge$_{2.85}$ ($\delta T_{Th}$ = 0.20) [19] and Nd$_2$PtGe$_3$ ($\delta T_{Th}$ = 0.12) [51].

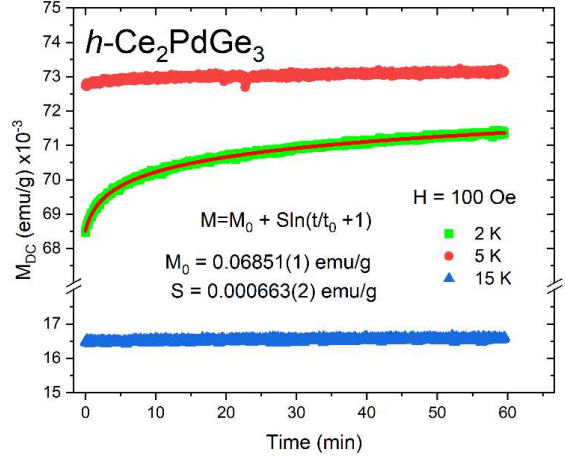

**Figure 7.** Time dependent remnant magnetization $M_{DC}(t)$ for $h$-Ce$_2$PdGe$_3$ Solid line represents fit to equation $M(t) = M_0 + S\ln(t/t_0 + 1)$.

One of the crucial features of the glassy like materials is the aging effect, which can be observed as a time evolution of the magnetization $M(t)$ under a constant applied field. $M(t)$ measured in the ZFC mode are presented in Fig. 7. It is clear that the DC magnetization increases with time at $T = 2$ K, while above the $T_f$(37 Hz) ($T = 5$ K and 15 K) the magnetization is almost time-independent (only a weak dependence is observed at $T = 5$ K). The non-equilibrium dynamical state below $T_f$ follows the equation [46]:

$$M(t) = M_0 + S\ln(t/t_0 + 1), \qquad (7)$$

where the temperature-dependent fitting parameters $M_0$ and $S$ are initial magnetization (magnetization at $t = 0$) and the magnetic viscosity, respectively. The $t_0$ parameter depends on measuring conditions and is usually a few orders of magnitude larger than $\tau_0$ [52]. The red line in Fig. 7 represents the best fit of above formula to the experimental data recorded at $T = 2$ K. The obtained values $M_0 = 0.00123(3)$ emu/g and $S = 0.000125(2)$ emu/g are comparable with Ho$_2$Pd$_{1.3}$Ge$_{2.7}$ [10], Dy$_2$Pd$_{1.25}$Ge$_{2.75}$ [18], Nd$_2$PtGe$_3$ [51]. The coexistence of an ageing effect and a shift in $\chi'(T)$ plot excludes a long range antiferromagnetic ordering and convincingly establishes $h$-Ce$_2$PdGe$_3$ as a cluster glass material.

An alternative approach to describing the relaxation phenomenon Is to use a stretched exponential law, expressed as a function [53], [54]:

$$M(t) = M_0 \pm M_g \, exp(-(t/\tau)^\beta), \qquad (8)$$

where $M_g$ described a glassy component of the magnetization and $\tau$ is a characteristic relaxation time. The best fit of the above expression for $h$-Ce$_2$PdGe$_3$ is presented in Fig. S5. The final fitting parameter - $\beta$, represents the stretching exponent. For various glassy systems, $\beta$ typically falls within the range $0 < \beta < 1$, where $\beta = 0$ indicates an absence of relaxation [55]. In the case of $h$-Ce$_2$PdGe$_3$, the value of this parameter is equal $\beta = 0.470(7)$, which is a relatively high value compared to the analogues known so far (e.g. $\beta = 0.233(4)$ for Tb$_2$Pt$_{1.2}$Ge$_{2.8}$). Probably, it indicates stronger relaxation process in this material. The discrepancy between the initial magnetization values calculated from the equation 7 (M₀ =

0.06851(1) emu/g) and the equation 8 (M₀ = 0.0721(3) emu/g) arises from a so called magnetic memory effect, which was discussed in detail in our previous article [19].

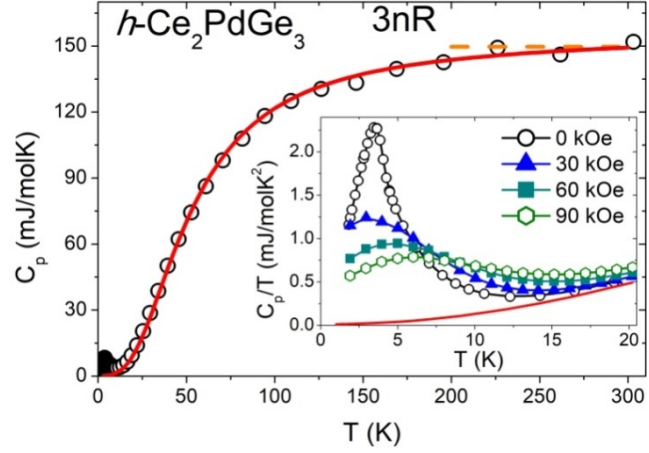

**Figure 7.** Temperature dependence of the heat capacity ($C_p$) for $h$-Ce$_2$PdGe$_3$ Dashed line represent Dulong-Petit limit (3$n$R). The red solid line represents the fit of the Debye formula. The inset shows plot of $C_p/T$ vs. $T$ at low temperatures measured for various applied magnetic fields.

The volume character of magnetic transition observed in $h$-Ce$_2$PdGe$_3$ was confirmed by specific heat measurements with and without an applied magnetic field. Gathered results are presented in Fig. 8. The $C_p(T)$ curve reaches a saturation at high temperatures, close to the expected value resulting from the Dulong-Petit law [56]: $3nR \approx 150$ J mol$^{-1}$ K$^{-1}$, where $n$ is the number of atoms per formula unit ($n = 6$) and R is the gas constant (R = 8.314 J mol$^{-1}$ K$^{-1}$). The occurrence of a broad hump near $T_f$(37 Hz) suggests a glassy transition rather than an antiferromagnetic ordering, which is in contrast to a sharp $C_p(T)$ peak in case of $t$-Ce$_2$PdGe$_3$ [23]. This behavior is better seen in the inset of Fig. 8, which displays $C_p/T$ in a function of temperature for different values of the external magnetic field. The observed peak is strongly affected by applied magnetic field - as the magnetic field strength increases, the peak gets blurrier and shifts toward higher temperatures, which is typical for spin-glass like compounds [57]. Moreover, a similar phenomenon was previously reported for many members of $RE_2TM$Ge$_3$ family, which exhibited a lack of long range magnetic ordering. We fitted the $C_p(T)$ with a model consisting of two components: electron specific heat and phonon specific heat. The former is represented by the Sommerfeld model and the latter by the Debye model, which can be written in the form of the following equation:

$$C_p(T) = \gamma T + 9nR \left(\frac{T}{\Theta_D}\right)^3 \int_0^{\Theta_D/T} \frac{x^4 \exp(x)}{[\exp(x)-1]^2} dx, \qquad (9)$$

where $\gamma$ is the electronic specific heat coefficient, $\Theta_D$ are the Debye temperature. In order to omit the influence of low-temperature anomalies associated with magnetic interactions, the fit was carried out from 15 K to 300 K and provided values of $\gamma = 11(3)$ mJmol$^{-1}$K$^{-2}$ and $\Theta_D = 213(2)$ K. The electronic density of states (DOS) at the Fermi surface ($E_F$) calculated by the formula [56]: $\gamma = k_B^2 \pi^4 N(E_F)/3$, is equal to 5(1) states/(eV atom).

The magnetocaloric effect (MCE) refers to the magnetothermal phenomenon observed in magnetic solids, where the temperature of the material increases or decreases adiabatically when a magnetic field is applied or removed, respectively [58]. This effect has significant technological implications, particularly in the development of environmentally friendly and energy-efficient refrigeration systems for various applications [59]. Furthermore, the scope of the magnetocaloric effect extends beyond its technological applications. It also provides valuable insights into the dominant magnetic interactions within a material and the potential competition among them. By studying the MCE, one can gain a deeper understanding of the magnetic properties and behavior of different materials. In addition, for materials showing magnetic frustration, it is possible to increase the magnetocaloric response [47], [60], [61]. Therefore, we decided to study MCE in $h$-Ce$_2$PdGe$_3$. The temperature dependence of the values of the parameters describing the MCE is shown in Fig. 9 and was determined from magnetic measurements. Magnetic entropy change $\Delta S_\mathrm{m}$ values were calculated from Maxwell's equation [58]:

$$\Delta S_\mathrm{m}(T, \mu_0 H) = \int_0^{\mu_0 H} (\delta M/\delta T) dH. \qquad (10)$$

For all studied conditions, $\Delta S_\mathrm{m}(T)$ is negative and forms a broad asymmetric peak located at ~7 K (i.e. above $T_\mathrm{f}$ in paramagnetic region) (Fig. 8(a)). The maximum value of $-\Delta S_\mathrm{m}$ is 0.8(1), 2.6(1), and 4.9(1) J kg$^{-1}$ K$^{-1}$ for a magnetic field change 20, 50, and 90 kOe, respectively. The maximum value of $-\Delta S_\mathrm{m}$ is observable over a fairly wide temperature range (between 5 and 8 K). In the literature, such $\Delta S_\mathrm{m}(T)$ behavior is called table-like MCE [58], [62]. Materials exhibiting a table-like MCE may be beneficial in the development of more efficient and controlled magnetic refrigeration systems, offering potential applications in energy-efficient cooling and refrigeration processes. The relationship of $\Delta S_\mathrm{m}$ vs $\Delta H$ for several temperatures is shown in panel (b) of Fig. 9. The $\Delta S_\mathrm{m}$ vs $\Delta H$ can be described by a power law: $\Delta S_\mathrm{m} = A \Delta H^w$ [63]. The determined value of the parameter $w$ for $T$ = 6.5 K is 1.13(2) and it is different from the value predicted by the theoretical model ($w$ = 2/3). The difference can be due to a number of reasons (e.g. this may be related to the emergence of short range magnetic interaction in paramagnetic region), and such a deviation in spin-glass-like materials is often observed [64].

Another parameter that determines the application potential of a given material is the relative cooling power (RCP), its value can be determined from the expression [58]:

$$\mathrm{RCP} = |\Delta S_\mathrm{m}| \delta T_\mathrm{FWHM}, \qquad (11)$$

where $|\Delta S_\mathrm{m}^\mathrm{MAX}|$ is the absolute maximum $\Delta S_\mathrm{m}$ value and $\delta T_\mathrm{FWHM}$ is the half-width of the $\Delta S_\mathrm{m}(T)$ peak. For a magnetic field change of 90 kOe, the RCP value is 63(3) Jkg$^{-1}$, and this is not a record high value. For the best materials, the value can reach 1000 Jkg$^{-1}$ for a magnetic field change of 50 kOe[58]. Maximum values of RCP as a function of the external magnetic field change are plotted in Fig. 9(c). This relationship can also be described by RCP = B $\Delta H^{(1+1/\delta)}$ exponential function. Determined from the fit, the value of the $\delta$ is 2.7(1) and is slightly lower than that predicted by mean-field theory ($\delta$ = 3) [65].

Additionally, using data from magnetic and specific heat measurements adiabatic temperature change $\Delta T_{ad}$ was estimated by using the following equation [58]:

$$\Delta T_{ad} = \frac{T}{c_p}\Delta S_m . \qquad (12)$$

The maximum values of $\Delta T_{ad}$ for $\Delta H = 50$ kOe is 8.1(1) K at ~9 K.

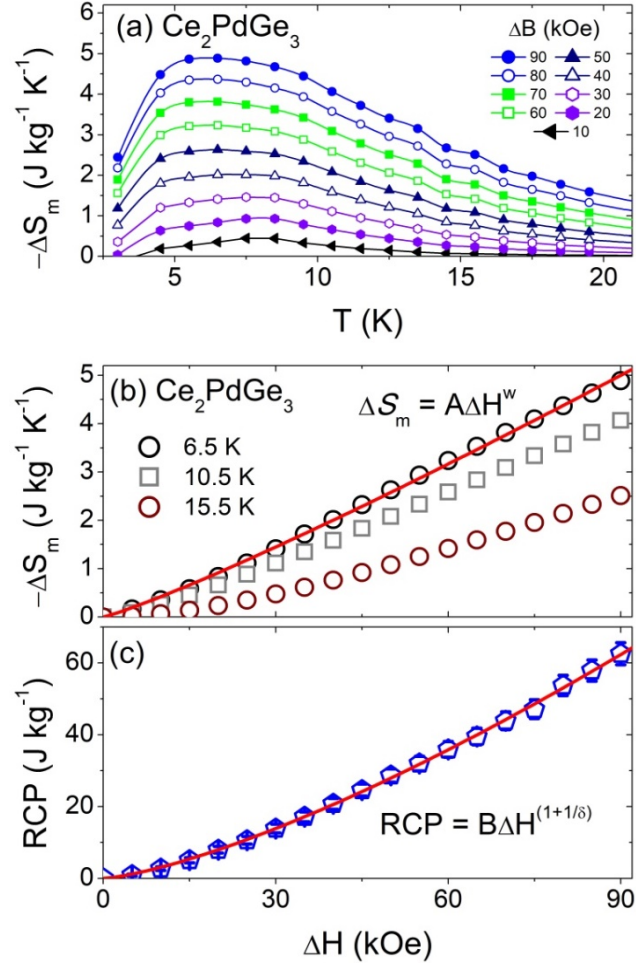

**Figure 8.** Magnetocaloric effect in $h$-$Ce_2PdGe_3$. (a) Magnetic entropy change $\Delta S_m$ obtained for various magnetic fields change. (b) $\Delta S_m$ vs. magnetic field change $\Delta H$ for several temperatures. (c) RCP vs. magnetic field change $\Delta H$. The solid lines represent fits with the power law.

Materials based on Ce are not renowned for their magnetocaloric effect. This holds true also in the case of $Ce_2PdGe_3$, where the values of size-determining parameters are relatively small, especially when compared to these best known materials [66]. At the same time, they are comparable to the values obtained for other compounds based on Ce [67], [68], [69]. The application potential of $Ce_2PdGe_3$ is low is and no enhancement of the magnetocaloric effect forced by magnetic frustration is observed [61].

Electrical resistivity $\rho$ measurements reveal a metallic character of $h$-$Ce_2PdGe_3$ - the value of $\rho$ successively decreases as the temperature decreases (Fig. 9). These results are in line with the XPS results. A broad hump is seen at ca. 50-100 K. The origin of this anomaly may have various sources, e.g., short-range magnetic interactions, crystal field influence, however the hypothesis about short-range magnetic interactions is the most plausible. It should be also noted that a similar feature is also observed in other AlB$_2$-type materials [17], [32], [70].A small peak is observed at the glassy transition temperature for a curve measured at $H = 0$ Oe (inset of Fig. 10).However, it can be also observed a broad cusp, which do not correspond to a magnetic transition.This feature can be explained by the lack of a periodically ordered state, which leads to a significant contribution to the electron scattering even above the freezing point. The correlation of this transition with magnetic properties of $h$-Ce$_2$PdGe$_3$ is indicated by its response to the applied magnetic field, i.e. smearing of the hump with an increase of $H$. The low-temperature $\rho$ decreases with applied magnetic field (negative magnetoresistance) up to ca. 20 K, above which the effect becomes barely visible.

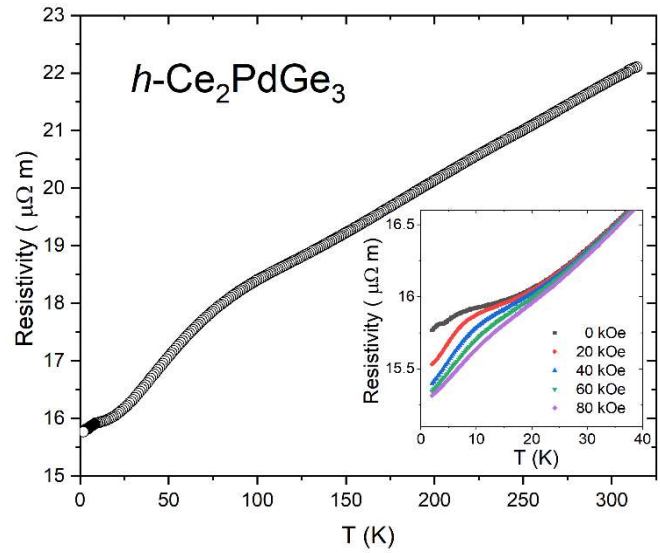

**Figure 8.**Electrical resistivity for $h$-Ce$_2$PdGe$_3$ measured at zero magnetic field. The inset shows the low temperature $\rho(T)$ dependence measured at different applied magnetic fields $H$.

Magnetoresistance measurements can also indicate spin disorder in the investigated compound [71]. At $T = 1.9$ K the magnetoresistance MR = 100%[$\rho$(90 kOe) - $\rho$(0 Oe)]/ $\rho$(0 Oe) = 2.9 %. The value of the residual resistivity ratio (RRR) was estimated employing the equation: RRR = $\rho$(300K)/$\rho$(2K) ≈ 1.4. This results lies in range usually expected for polycrystalline samples with internal defects or disorder.

Table 3 shows the comparison of reported physical properties of Ce-bearing AlB$_2$ and α-ThSi$_2$ type compounds. It can be observed that for all of them the effective magnetic moment is close (taking into account measurement uncertainties) to the theoretical value of the effective magnetic moment Ce$^{3+}$, which suggests that in all reported compounds there are magnets with a local moment of 4$f$ with no or negligible contribution of transition metal atoms in total magnetic moment. Most of the reported materials have both hexagonal and tetragonal variants known. They are stabilized either by a slight change of the *TM*:*X* ratio or by the different synthetic route[10], [15], [18], [19], [48], [53], [72], [73], [74], [75], [76], [77], [78]. It is known that there are two general factors crucial for the thermodynamic stability of the AlB$_2$ crystal structure: the relative sizes of atoms at the 1$a$ and 2$d$ sites and the valence electron count. The metastable variant, however, may be kinetically favored, which explains the importance of the heat treatment process. The structural variants differ significantly in their magnetic properties.

**Table 2.** Summary of physical properties of $h$-Ce$_2$PdGe$_3$. The values of magnetocaloric parameters are collected for a magnetic field change value $\Delta H$ = 50 kOe. Numbers in parentheses are statistical uncertainties of fitted parameters. The total error is expected to be larger due to experimental factors (e.g. trace amounts of impurity phases).

| Parameter | Ce$_2$PdGe$_3$ ($P6/mmm$) |
|---|---|
| $T_{irr}$ (K) | 3.1(1) |
| $T_f$(37 Hz)(K) | 3.44(2) |
| $\phi$ | 1.2(1) |
| $\delta T_f$ | 0.012(2) |
| $E_a/k_B$ (K) | 5.94(6) |
| $T_0$ (K) | 3.10(3) |
| $\tau_0$ (s) | 10$^{-9}$ |
| $\delta T_{Th}$ | 0.10(2) |
| $\theta_{CW}$ (K) | -3.7(1) |
| $\mu_{eff}$ ($\mu_B$) | 2.58(8) |
| $M_0$ (emu/g) × 10$^{-3}$ | 1.23(7) |
| $S$ (emu/g) × 10$^{-4}$ | 1.25(2) |
| RRR | 1.4(1) |
| $\gamma$ (mJmol$^{-1}$K$^{-2}$) | 11(3) |
| $\Theta_D$ (K) | 213(2) |
| $\Delta S_m$ (J kg$^{-1}$ K$^{-1}$) | 2.6(1) |
| RCP (J kg$^{-1}$) | 29(2) |
| $\Delta T_{ad}$ (K) | 8.1(1) |

**Table 2.** Comparison of lattice constants and physical properties of ternary Ce-bearing AlB$_2$ and $\alpha$-ThSi$_2$ type compounds. Cell volumes for AlB$_2$-type compounds are multiplied by 4 to facilitate the comparison of relative volumes of allotropic forms.

| Compound | Crystal structure | Lattice parameters (Å) | Cell volume (Å$^3$) | Magnetic properties | $T_t$ (K) | $\mu_{eff}$ ($\mu_B$) | $\theta_{CW}$ (K) | Ref. |
|---|---|---|---|---|---|---|---|---|
| $\alpha$-Ce$_2$RhGe$_3$ | $\alpha$-ThSi$_2$ ($I4_1/amd$) | $a$ = 4.2034(6) $c$ = 14.770(3) | 260.96 | Spin glass like | 3.6 | 2.7 | -30.7 | [22] |
| $\beta$-Ce$_2$RhGe$_3$ | AlB$_2$ ($P6/mmm$) | $a$ = 4.2615(7) $c$ = 4.1813(9) | 263.04 | Spin glass like | 12 | 2.52 | -20.85 | [22] |
| CeAl$_{1.2}$Si$_{0.8}$ | $\alpha$-ThSi$_2$ ($I4_1/amd$) | $a$ = 4.280 $c$ = 14.755 | 270.29 | ferromagnet | 4.2 | 2.64 | -11.0 | [79] |
| CeAl$_{1.5}$Si$_{0.5}$ | $\alpha$-ThSi$_2$ ($I4_1/amd$) | $a$ = 4.280 $c$ = 14.900 | 272.80 | ferromagnet | 6.35 | 2.69 | -16.2 | [79] |
| CeAl$_{1.2}$Ge$_{0.8}$ | $\alpha$-ThSi$_2$ ($I4_1/amd$) | $a$ = 4.298 $c$ = 14.826 | 273.88 | antiferromagnet | 3.4 | 2.65 | -16.5 | [79] |
| CeAl$_{1.5}$Ge$_{0.5}$ | AlB$_2$ ($P6/mmm$) | $a$ = 4.354 $c$ = 4.342 | 213.60 | ferromagnet | 6.15 | 2.63 | -45.0 | [79] |
| CeAu$_{0.75}$Ge$_{1.25}$ | AlB$_2$ ($P6/mmm$) | $a$ = 4.335(1) $c$ = 4.226(1) | 206.43 | ferromagnet | 6.0(5) | 2.51(3) | 12(2) | [80] |
| CePd$_{0.63}$Ge$_{1.37}$ | AlB$_2$ ($P6/mmm$) | $a$ = 4.289 (3) $c$ = 4.183(3) | 266.56 | ferromagnet | 3.0(5) | 2.51(2) | 2(2) | [80] |
| $h$-Ce$_2$PdGe$_3$ | AlB$_2$ ($P6/mmm$) | $a$ = 4.2602(2) $c$ = 4.2511(1) | 267.27 | Cluster glass | 3.1 | 2.56 | -3.66(8) | This work |
| t-Ce$_2$PdGe$_3$ | α-ThSi$_2$ (I4$_1$/amd) | a = 4.24440(8) c = 14.7928(2) | 266.49 | Antiferromagnet (with 2 transitions) | 10.7/9.6 | 2.51 | -6.3 | [23], [24] |
| CeCu$_{0.6}$Ge$_{1.4}$ | AlB$_2$ (P6/mmm) | a = 4.289(3) c = 4.183(3) | 199.92 | - | - | - | - | [81] |
| CeCu$_{0.4}$Ge$_{1.6}$ | α-ThSi$_2$ (I4$_1$/amd) | a = 4.246 c = 14.494 | 261.31 | - | - | - | - | [81] |

## 4. Conclusion

A polycrystalline sample of a hexagonal Ce$_2$PdGe$_3$ (space group P6/mmm, no. 191) was synthetized using an arc melting technique. Detailed studies of AC and DC susceptibility exhibit that this material can be classified as a cluster glass with the freezing temperature $T_f$(37 Hz) = 3.44 K. It is in line with the heat capacity and resistivity measurements. The observed cluster glass like behavior results from the coexistence of a disordered structure and magnetic frustration, which are characteristic features of many RE$_2$TMGe$_3$ compounds. The polymorphism observed in the system is also reported for other RE$_2$TMGe$_3$ and RE$_2$TMSi$_3$ materials, highlighting the comparable stability of the AlB$_2$ and α-ThSi$_2$ structure types. The XPS analyses validate the high quality of the Ce$_2$PdGe$_3$ specimen, revealing no indications of oxide development and demonstrating that Ce ions are primarily in the stable 3+ state. The obtained values of the magnetocaloric parameters for h-Ce$_2$PdGe$_3$ are not high, and noimprovement of the MCE by the occurrence of magnetic frustration was observed. The $\Delta S_m(T)$ forms a wide flat peak, i.e. it exhibits the so-called table-like MCE, where the value of $\Delta S_m$ is practically constant over a wide temperature range, and this positively affects the performance of such magnetocaloric material.

**Supporting Information**
Supporting Information is available from the Wiley Online Library or from the author.

**Acknowledgements**
This work was supported by National Science Centre Poland under project 2023/07/X/ST11/00345("MINIATURA"). K.S. acknowledge the financial support of the National Science Centre Poland under project2021/41/B/ST5/02894 ("OPUS").

**Availability of data**
The data that support the findings of this study are openly available in Zenodo at https://doi.org/10.5281/zenodo.15090260, reference number 15090260.

# References

[1] S. Paschen and Q. Si, "Quantum phases driven by strong correlations," *Nature Reviews Physics*, vol. 3, no. 1, pp. 9–26, 2021, doi: 10.1038/s42254-020-00262-6.

[2] Z. F. Weng, M. Smidman, L. Jiao, X. Lu, and H. Q. Yuan, "Multiple quantum phase transitions and superconductivity in Ce-based heavy fermions," *Reports on Progress in Physics*, vol. 79, no. 9, p. 94503, Aug. 2016, doi: 10.1088/0034-4885/79/9/094503.

[3] P. Coleman and A. J. Schofield, "Quantum criticality," *Nature*, vol. 433, no. 7023, pp. 226–229, 2005, doi: 10.1038/nature03279.

[4] F Steglich et al., "Quantum critical phenomena in undoped heavy-fermion metals," *Journal of Physics: Condensed Matter*, vol. 8, no. 48, p. 9909, 1996, doi: 10.1088/0953-8984/8/48/016.

[5] J. Dshemuchadse and W. Steurer, "Some Statistics on Intermetallic Compounds," *Inorg Chem*, vol. 54, no. 3, pp. 1120–1128, Feb. 2015, doi: 10.1021/ic5024482.

[6] R.-D. Hoffmann and R. Pöttgen, "AlB2-related intermetallic compounds – a comprehensive view based on group-subgroup relations," *Z KristallogrCryst Mater*, vol. 216, no. 3, pp. 127–145, 2001, doi: https://doi.org/10.1524/zkri.216.3.127.20327.

[7] C. Zheng and R. Hoffmann, "Conjugation in the 3-connected net: the aluminum diboride and thorium disilicide structures and their transition-metal derivatives," *Inorg Chem*, vol. 28, no. 6, pp. 1074–1080, Mar. 1989, doi: 10.1021/ic00305a015.

[8] M. Nentwich et al., "Structure variations within RSi2 and R2TSi3 silicides. Part I. Structure overview," *Acta Crystallographica Section B*, vol. 76, pp. 177–200, Mar. 2020, doi: 10.1107/S2052520620001043.

[9] M. Nentwich et al., "Structure variations within RSi 2 and R 2Si 3 silicides. Part II. Structure driving factors," *Acta Crystallographica Section B*, vol. 76, no. 3, pp. 378–410, Jun. 2020, doi: 10.1107/S2052520620003893.

[10] L. S. Litzbarski, T. Klimczuk, and M. J. Winiarski, "Ho2Pd1.3Ge2.7 – a ternary AlB2-type cluster glass system," *RSC Adv.*, vol. 11, no. 41, pp. 25187–25193, 2021, doi: 10.1039/D1RA04422B.

[11] A. Brown, "MX 2 compounds of thorium and the polymorphism of thorium disilicide," *Acta Crystallogr*, vol. 14, no. 8, pp. 860–865, Aug. 1961, doi: 10.1107/S0365110X61002497.

[12] A. BROWN and J. J. NORREYS, "Beta-Polymorphs of Uranium and Thorium Disilicides," *Nature*, vol. 183, no. 4662, p. 673, 1959, doi: 10.1038/183673a0.

[13] J. Dshemuchadse and W. Steurer, "More statistics on intermetallic compounds – ternary phases," *Acta Crystallographica Section A*, vol. 71, pp. 335–345, Apr. 2015, doi: 10.1107/S2053273315004064.

[14] M. Szlawska, "Spin-glass freezing in single-crystalline Pr2NiSi3," *Intermetallics (Barking)*, vol. 115, p. 106616, 2019, doi: https://doi.org/10.1016/j.intermet.2019.106616.

[15] L. S. Litzbarski, M. Łapiński, T. Klimczuk, and M. J. Winiarski, "Synthesis and physical properties of Sm2PdGe3 in a context of RE2PdGe3 family," *J Solid State Chem*, vol. 340, p. 125010, 2024, doi: https://doi.org/10.1016/j.jssc.2024.125010.

[16] T. Klimczuk et al., "Crystal structure and physical properties of new Ca2TGe3 (T = Pd and Pt) germanides," *J Solid State Chem*, vol. 243, Aug. 2016, doi: 10.1016/j.jssc.2016.07.029.

[17] S. Sarkar et al., "Synthetically tuned structural variations in CePdxGe2-x (x = 0.21, 0.32, 0.69) towards diverse physical properties," *Inorg Chem Front*, vol. 4, no. 2, pp. 241–255, 2017, doi: 10.1039/c6qi00366d.

[18] L. Litzbarski, T. Klimczuk, and M. Winiarski, "Synthesis, structure and physical properties of new intermetallic spin glass-like compounds RE2PdGe3 (RE = Tb and Dy)," *Journal of Physics: Condensed Matter*, vol. 32, Feb. 2020, doi: 10.1088/1361-648X/ab73a4.

[19] L. S. Litzbarski *et al.*, "Cluster-spin-glass behavior in new ternary RE2PtGe3 compounds (RE = Tb, Dy, Ho)," *Mater Res Express*, 2024, [Online]. Available: http://iopscience.iop.org/article/10.1088/2053-1591/ad7444

[20] S. Sarkar and S. C. Peter, "Structural phase transitions in a new compound Eu2AgGe 3," *Inorg Chem*, vol. 52, no. 17, pp. 9741–9748, 2013, doi: 10.1021/ic400369a.

[21] A. M. Strydom, "Structure determination of the new rare-earth compound Ce2PdGe3 : research in action," *S Afr J Sci*, vol. 99, pp. 419–421, 2003.

[22] D. Kalsi, U. Subbarao, S. Rayaprol, and S. C. Peter, "Structural and magnetic properties in the polymorphs of CeRh0.5Ge1.5," *J Solid State Chem*, vol. 212, pp. 73–80, 2014, doi: https://doi.org/10.1016/j.jssc.2014.01.015.

[23] R. Baumbach *et al.*, "Complex magnetism and strong electronic correlations in Ce2PdGe3," *Phys Rev B*, vol. 91, Aug. 2014, doi: 10.1103/PhysRevB.91.035102.

[24] J. Kitagawa, Y. Muro, N. Takeda, and M. Ishikawa, "Low-Temperature Magnetic Properties of Several Compounds in Ce-Pd-X (X=Si, Ge and Al) Ternary Systems," *J Physical Soc Japan*, vol. 66, no. 7, pp. 2163–2174, Jul. 1997, doi: 10.1143/JPSJ.66.2163.

[25] A. Bhattacharyya, C. Ritter, D. T. Adroja, F. C. Coomer, and A. M. Strydom, "Exploring the complex magnetic phase diagram of ${\mathrm{Ce}}_{2}{\mathrm{PdGe}}_{3}$: A neutron powder diffraction and $\ensuremath{\mu}\mathrm{SR}$ study," *Phys. Rev. B*, vol. 94, no. 1, p. 14418, 2016, doi: 10.1103/PhysRevB.94.014418.

[26] J. Rodriguez-Carvajal, "Recent Advances in Magnetic Structure Determination by Neutron Powder Diffraction," *Physica B Condens Matter*, vol. 192, pp. 55–69, Oct. 1993, doi: 10.1016/0921-4526(93)90108-I.

[27] S. Majumdar and E. V Sampathkumaran, "Observation of enhanced magnetic transition temperature in Nd2PdGe3 and superconductivity in Y2PdGe3," *Phys Rev B*, vol. 63, May 2001.

[28] S. Majumdar, M. M. Kumar, and E. V Sampathkumaran, "Magnetic behavior of a new compound, Gd2PdGe3," *J Alloys Compd*, vol. 288, no. 1, pp. 61–64, 1999, doi: https://doi.org/10.1016/S0925-8388(99)00119-X.

[29] A. P. Ramirez, "Strongly Geometrically Frustrated Magnets," *Annual Review of Materials Science*, vol. 24, no. 1, pp. 453–480, Aug. 1994, doi: 10.1146/annurev.ms.24.080194.002321.

[30] Y. Li, "YbMgGaO4: A Triangular-Lattice Quantum Spin Liquid Candidate," *Adv Quantum Technol*, vol. 2, no. 12, p. 1900089, 2019, doi: https://doi.org/10.1002/qute.201900089.

[31] B. Schmidt, J. Sichelschmidt, K. M. Ranjith, Th. Doert, and M. Baenitz, "Yb delafossites: Unique exchange frustration of $4f$ spin-$\frac{1}{2}$ moments on a perfect triangular lattice," *Phys. Rev. B*, vol. 103, no. 21, p. 214445, 2021, doi: 10.1103/PhysRevB.103.214445.

[32] P. Zhi-Yan, C. Chong-De, B. Xiao-Jun, S. Rui-Bo, Z. Jian-Bang, and D. Li-Bing, "Structures and physical properties of R2TX3 compounds," *Chinese Physics B*, vol. 22, p. 56102, May 2013, doi: 10.1088/1674-1056/22/5/056102.

[33] M. Brun, A. Berthet, and J. C. Bertolini, "XPS, AES and Auger parameter of Pd and PdO," *J Electron SpectrosRelat Phenomena*, vol. 104, no. 1, pp. 55–60, 1999, doi: https://doi.org/10.1016/S0368-2048(98)00312-0.

[34] L. Liu, "Effects of Spin-Orbit Coupling in Si and Ge," *Physical Review*, vol. 126, no. 4, pp. 1317–1328, May 1962, doi: 10.1103/PhysRev.126.1317.

[35] O. Gunnarsson and K. Schönhammer, "Electron spectroscopies for Ce compounds in the impurity model," *Phys Rev B*, vol. 28, no. 8, pp. 4315–4341, Oct. 1983, doi: 10.1103/PhysRevB.28.4315.

[36] P. Skokowski, K. Synoradzki, M. Werwiński, A. Bajorek, G. Chełkowska, and T. Toliński, "Electronic structure of CeCo1−xFexGe3 studied by X-ray photoelectron spectroscopy and first-principles calculations," *J Alloys Compd*, vol. 787, pp. 744–750, 2019, doi: https://doi.org/10.1016/j.jallcom.2019.02.056.

[37] A. Szajek, G. Chełkowska, T. Toliński, and A. Kowalczyk, "Electronic structure of CeCu4In from band structure calculations and X-ray photoelectron spectroscopy," *J Magn Magn Mater*, vol. 596, p. 171938, 2024, doi: https://doi.org/10.1016/j.jmmm.2024.171938.

[38] J. Goraus, G. Chełkowska, A. Kowalczyk, and M. Falkowski, "A combined first-principles calculations and X-ray photoelectron spectroscopy of Ce2T3X9 (T = Rh, Ru, Ir; X = Al, Ga): Possible strong topological insulator state in Ce2Ir3Al9," *ComputMater Sci*, vol. 231, p. 112586, 2024, doi: https://doi.org/10.1016/j.commatsci.2023.112586.

[39] S. Datta *et al.*, "Layer-resolved electronic behavior in a Kondo lattice system, CeAgAs2," *Journal of Physics: Condensed Matter*, vol. 35, no. 23, p. 235601, 2023, doi: 10.1088/1361-648X/acc5c9.

[40] P. Skokowski, K. Synoradzki, M. Werwiński, T. Toliński, A. Bajorek, and G. Chełkowska, "Influence of Pr substitution on the physical properties of the ${\mathrm{Ce}}_{1\ensuremath{-}x}{\mathrm{Pr}}_{x}\mathrm{Co}{\mathrm{Ge}}_{3}$ system: Combined experimental and first-principles study," *Phys Rev B*, vol. 102, no. 24, p. 245127, Dec. 2020, doi: 10.1103/PhysRevB.102.245127.

[41] S. Pandey *et al.*, "Intermediate valence and spin fluctuations near a quantum critical point in $\mathrm{Ce}{\mathrm{Ru}}_{2\text{\ensuremath{-}}x}{\mathrm{Co}}_{x}{\mathrm{Ge}}_{2}$," *Phys Rev B*, vol. 108, no. 1, p. 14407, Jul. 2023, doi: 10.1103/PhysRevB.108.014407.

[42] J. Jensen and A. Mackintosh, *Rare Earth Magnetism: Structures and Excitations*. 1991.

[43] A. P. Ramirez, "Thermodynamic measurements on geometrically frustrated magnets (invited)," *J Appl Phys*, vol. 70, no. 10, pp. 5952–5955, Nov. 1991, doi: 10.1063/1.350088.

[44] J. A. Mydosh, *Spin glasses: an experimental introduction*. Taylor and Francis, 1993.

[45] S. Chakraborty *et al.*, "Ground-state degeneracy and complex magnetism of geometrically frustrated Gd2 Ir0.97 Si2.97," *Phys Rev B*, vol. 106, no. 22, Dec. 2022, doi: 10.1103/PhysRevB.106.224427.

[46] J. A. Mydosh, "Spin glasses: redux: an updated experimental/materials survey," *Reports on Progress in Physics*, vol. 78, no. 5, p. 52501, 2015, doi: 10.1088/0034-4885/78/5/052501.

[47] S. Pakhira, C. Mazumdar, R. Ranganathan, S. Giri, and M. Avdeev, "Large magnetic cooling power involving frustrated antiferromagnetic spin-glass state in R2NiSi3 (R=Gd,Er)," *Phys Rev B*, vol. 94, no. 10, p. 104414, Sep. 2016, doi: 10.1103/PhysRevB.94.104414.

[48] S. Pakhira, C. Mazumdar, R. Ranganathan, and S. Giri, "Chemical disorder driven reentrant spin cluster glass state formation and associated magnetocaloric properties of Nd2Ni0.94Si2.94," *Phys. Chem. Chem. Phys.*, vol. 20, no. 10, pp. 7082–7092, 2018, doi: 10.1039/C7CP08574E.

[49] Y. G. Joh, R. Orbach, G. G. Wood, J. Hammann, and E. Vincent, "Extraction of the Spin Glass Correlation Length," *Phys. Rev. Lett.*, vol. 82, no. 2, pp. 438–441, 1999, doi: 10.1103/PhysRevLett.82.438.

[50] J. Souletie and J. L. Tholence, "Critical slowing down in spin glasses and other glasses: Fulcher versus power law.," *Phys Rev B Condens Matter*, vol. 32, no. 1, pp. 516–519, Jul. 1985, doi: 10.1103/physrevb.32.516.

[51] L. S. Litzbarski, M. J. Winiarski, P. Skokowski, T. Klimczuk, and B. Andrzejewski, "Investigation of magnetic order in a new intermetallic compound Nd2PtGe3," *J Magn Magn Mater*, vol. 521, p. 167494, 2021, doi: https://doi.org/10.1016/j.jmmm.2020.167494.

[52] H. Nair *et al.*, "Memory effect in Dy 0.5 Sr 0.5 MnO 3 single crystals," *J Phys Condens Matter*, vol. 22, p. 346002, Sep. 2010, doi: 10.1088/0953-8984/22/34/346002.

[53] S. Pakhira, C. Mazumdar, R. Ranganathan, and S. Giri, "Magnetic phase inhomogeneity in frustrated intermetallic compound Sm 2 Ni 0.87 Si 2.87," *J Alloys Compd*, vol. 742, Jan. 2018, doi: 10.1016/j.jallcom.2018.01.145.

[54] J. Kroder *et al.*, "Spin glass behavior in the disordered half-Heusler compound IrMnGa," *Phys. Rev. B*, vol. 99, no. 17, p. 174410, 2019, doi: 10.1103/PhysRevB.99.174410.

[55] S. Pakhira, N. S. Sangeetha, V. Smetana, A.-V. Mudring, and D. C. Johnston, "Ferromagnetic cluster-glass phase in $\mathrm{Ca}{({\mathrm{Co}}_{1\ensuremath{-}x}{\mathrm{Ir}}_{x})}_{2\ensuremath{-}y}{\mathrm{As}}_{2}$ crystals," *Phys. Rev. B*, vol. 102, no. 2, p. 24410, 2020, doi: 10.1103/PhysRevB.102.024410.

[56] A. F. Tarı, "The Specific heat of matter at low temperatures," 2003. [Online]. Available: https://api.semanticscholar.org/CorpusID:92908264

[57] C. Tien, C. Feng, C. Wur, and J. Lu, "Ce2CuGe3: A nonmagnetic atom-disorder spin glass," *Phys Rev B*, vol. 61, pp. 12158–74065, May 2000, doi: 10.1103/PhysRevB.61.12151.

[58] V. Franco, J. S. Blázquez, J. J. Ipus, J. Y. Law, L. M. Moreno-Ramírez, and A. Conde, "Magnetocaloric effect: From materials research to refrigeration devices," *Prog Mater Sci*, vol. 93, pp. 112–232, 2018, doi: https://doi.org/10.1016/j.pmatsci.2017.10.005.

[59] V. K. Pecharsky and K. A. Gschneidner Jr, "Magnetocaloric effect and magnetic refrigeration," *J Magn Magn Mater*, vol. 200, no. 1, pp. 44–56, 1999, doi: https://doi.org/10.1016/S0304-8853(99)00397-2.

[60] K. Synoradzki, "Magnetocaloric effect in spin-glass-like GdCu4Mn compound," *J Magn Magn Mater*, vol. 546, p. 168857, 2022, doi: https://doi.org/10.1016/j.jmmm.2021.168857.

[61] M. E. Zhitomirsky, "Enhanced magnetocaloric effect in frustrated magnets," *Phys Rev B*, vol. 67, no. 10, p. 104421, Mar. 2003, doi: 10.1103/PhysRevB.67.104421.

[62] L. Tian, Z. Mo, J. Gong, X. Gao, G. Liu, and J. Shen, "Large table-like magnetocaloric effect in boron-doped Er5Si3B0.5 compound," *J Appl Phys*, vol. 134, Aug. 2023, doi: 10.1063/5.0161680.

[63] V. Franco, J. Blázquez, and A. Conde, "Field dependence of the magnetocaloric effect in materials with a second order phase transition: A master curve for the magnetic entropy change," *Appl Phys Lett*, vol. 89, p. 222512, Oct. 2006, doi: 10.1063/1.2399361.

[64] K. Synoradzki, "Magnetocaloric effect in spin-glass-like GdCu4Mn compound," *J Magn Magn Mater*, vol. 546, p. 168857, 2022, doi: https://doi.org/10.1016/j.jmmm.2021.168857.

[65] V. Franco and A. Conde, "Scaling laws for the magnetocaloric effect in second order phase transitions: From physics to applications for the characterization of materials,"

*International Journal of Refrigeration*, vol. 33, no. 3, pp. 465–473, 2010, doi: https://doi.org/10.1016/j.ijrefrig.2009.12.019.

[66] W. Liu *et al.*, "A study on rare-earth Laves phases for magnetocaloric liquefaction of hydrogen," *Appl Mater Today*, vol. 29, p. 101624, 2022, doi: https://doi.org/10.1016/j.apmt.2022.101624.

[67] K. Synoradzki *et al.*, "Ferromagnetic CeSi1.2Ga0.8 alloy: Study on magnetocaloric and thermoelectric properties," *J Magn Magn Mater*, vol. 547, p. 168833, 2022, doi: https://doi.org/10.1016/j.jmmm.2021.168833.

[68] K. Synoradzki, P. Skokowski, Ł. Frąckowiak, M. Koterlyn, and T. Toliński, "Magnetocaloric properties in cryogenic temperature range of ferromagnetic CeSi1.3Ga0.7 alloy," *J Magn Magn Mater*, vol. 547, p. 168886, 2022, doi: https://doi.org/10.1016/j.jmmm.2021.168886.

[69] K. Synoradzki, D. Das, A. Frąckowiak, D. Szymański, P. Skokowski, and D. Kaczorowski, "Study on magnetocaloric and thermoelectric application potential of ferromagnetic compound CeCrGe3," *J Appl Phys*, vol. 126, no. 7, p. 075114, Aug. 2019, doi: 10.1063/1.5107450.

[70] D. X. Li, A. Kimura, Y. Haga, S. Nimori, and T. Shikama, "Magnetic anisotropy and spin-glass behavior in single crystalline U2PdSi3," *Journal of Physics: Condensed Matter*, vol. 23, no. 7, p. 076003, 2011, doi: 10.1088/0953-8984/23/7/076003.

[71] M. Roman, L. Litzbarski, T. Klimczuk, and K. K. Kolincio, "Crossover from charge density wave stabilized antiferromagnetism to superconductivity in ${\mathrm{Nd}}_{1\ensuremath{-}x}{\mathrm{La}}_{x}{\mathrm{NiC}}_{2}$ compounds," *Phys. Rev. B*, vol. 99, no. 24, p. 245152, 2019, doi: 10.1103/PhysRevB.99.245152.

[72] H. Świątek, S. Królak, L. Litzbarski, I. Oshchapovsky, M. J. Winiarski, and T. Klimczuk, "Detailed studies of superconducting properties of Y2Pd1.25Ge2.75," *J Alloys Compd*, vol. 971, p. 172712, 2024, doi: https://doi.org/10.1016/j.jallcom.2023.172712.

[73] L. S. Litzbarski *et al.*, "Intermetallic disordered magnet ${\mathrm{Gd}}_{2}{\mathrm{Pt}}_{1.1}{\mathrm{Ge}}_{2.9}$ and its relation to other ${\mathrm{AlB}}_{2}$-type compounds," *Phys. Rev. B*, vol. 105, no. 5, p. 54427, 2022, doi: 10.1103/PhysRevB.105.054427.

[74] S. Pakhira, C. Mazumdar, and R. Ranganathan, "Magnetocaloric properties of cluster glass compound Pr2Ni0.95Si2.95," *Intermetallics (Barking)*, vol. 111, p. 106490, 2019, doi: https://doi.org/10.1016/j.intermet.2019.106490.

[75] S. Pakhira, C. Mazumdar, and R. Ranganathan, "Low-field induced large magnetocaloric effect in Tm2Ni0.93Si2.93: Influence of short-range magnetic correlation," *Journal of Physics: Condensed Matter*, vol. 29, Oct. 2017, doi: 10.1088/1361-648X/aa9736.

[76] S. Pakhira, A. K. Kundu, C. Mazumdar, and R. Ranganathan, "Role of random magnetic anisotropy on the valence, magnetocaloric and resistivity properties in a hexagonal Sm2Ni0.87Si2.87 compound," *Journal of Physics: Condensed Matter*, vol. 30, no. 21, p. 215601, 2018, doi: 10.1088/1361-648X/aabc22.

[77] S. Pakhira, C. Mazumdar, M. Avdeev, R. Bhowmik, and R. Ranganathan, "Spatially limited antiferromagnetic order in a cluster glass compound Tb2Ni0.90Si2.94," *J Alloys Compd*, vol. 785, May 2019, doi: 10.1016/j.jallcom.2019.01.123.

[78] M. Kundu, S. Pakhira, D. Paudyal, N. Lakshminarasimhan, R. Ranganathan, and C. Mazumdar, "Magnetocaloric response with significant mechanical efficiency in frustrated intermetallic compound Pr2Co0.86Si2.88," *Intermetallics (Barking)*, vol. 151, p. 107730, 2022, doi: https://doi.org/10.1016/j.intermet.2022.107730.

[79] H. Flandorfer *et al.*, "The Systems Ce–Al–(Si, Ge): Phase Equilibria and Physical Properties," *J Solid State Chem*, vol. 137, no. 2, pp. 191–205, 1998, doi: https://doi.org/10.1006/jssc.1997.7660.

[80] C. D. W. Jones, R. A. Gordon, F. J. DiSalvo, R. Pöttgen, and R. K. Kremer, "Magnetic behaviour of two AlB2-related germanides: CePd0.63Ge1.37 and CeAu0.75Ge1.25," *J Alloys Compd*, vol. 260, no. 1, pp. 50–55, 1997, doi: https://doi.org/10.1016/S0925-8388(97)00159-X.

[81] G. Nakamoto, T. Hagiuda, and M. Kurisu, "Formation of AlB 2- and ThSi 2-type structures in Ce(Cu x Ge y ) 2 system," *Physica B-condensed Matter - PHYSICA B*, vol. 312, pp. 277–279, Oct. 2002, doi: 10.1016/S0921-4526(01)01299-6.

# Supporting Information

**Cluster glass behavior and magnetocaloric effect in the hexagonal polymorph of disordered Ce$_2$PdGe$_3$**

*Leszek S. Litzbarski*[*], *Kamil Balcarek, Anna Bajorek, Tomasz Klimczuk, Michał J. Winiarski, Karol Synoradzki*

In the Supplementary Information (SI), graphs illustrating the results of measurements and analysis of XPS spectra for the Ce$_2$PdGe$_3$ compound were included. Details of the measurements are provided in the main text of the manuscript.

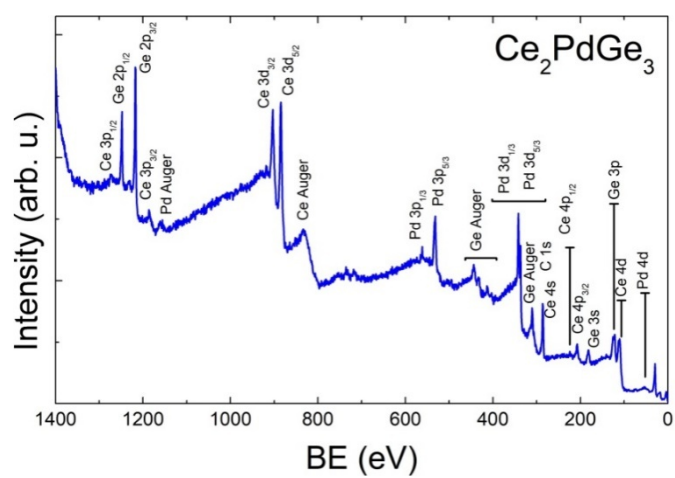

Fig. S1. XPS spectrum of Ce$_2$PdGe$_3$ collected for the 0–1400 eV binding energy range.

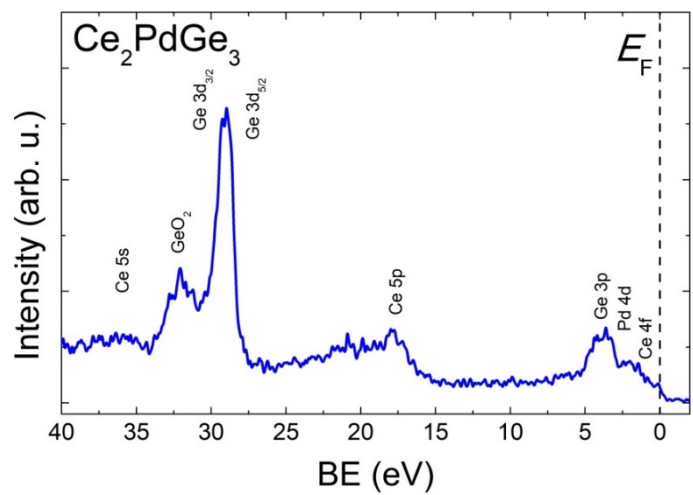

Fig. S2. XPS valence band spectrum of Ce$_2$PdGe$_3$.

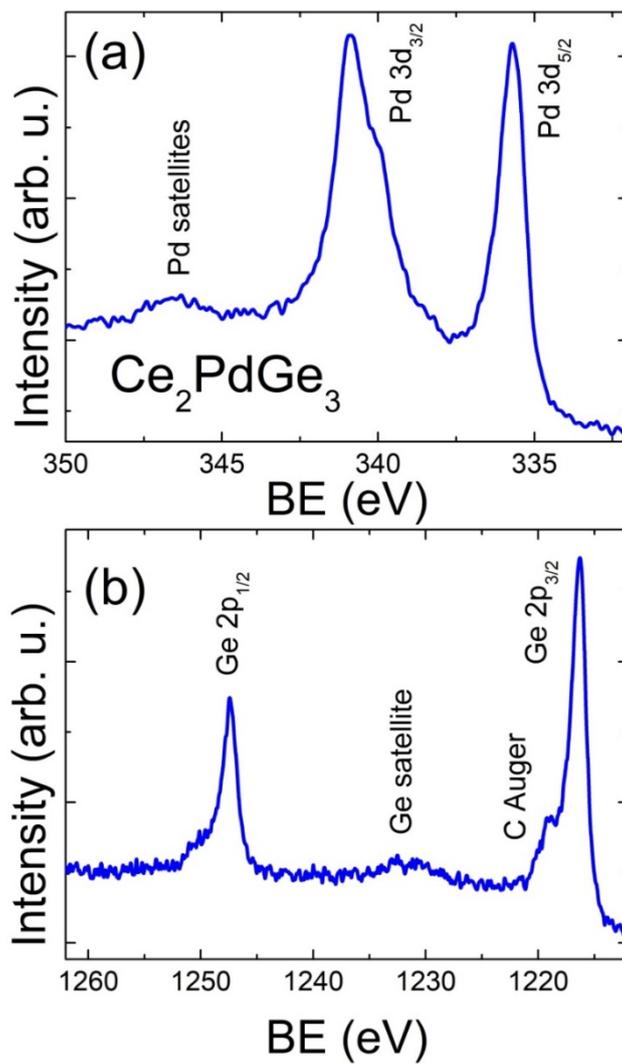

Fig. S3. The P*d* 3d (a) and the Ge 2*p* XPS spectra (b) of Ce$_2$PdGe$_3$.

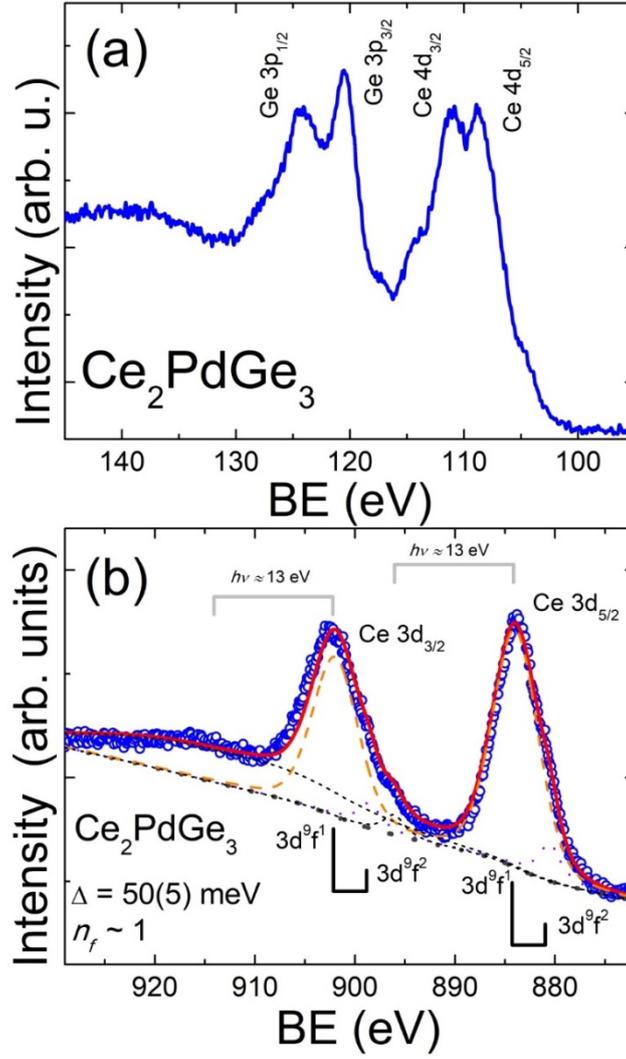

Fig. S4. The Ce 4$d$ (a) and the Ce 3$d$ (b) XPS spectra obtained for Ce$_2$PdGe$_3$. The satellite lines observed in the Ce 3d XPS spectra are interpreted as plasmon losses with an energy of $h\nu \approx 13$ eV.

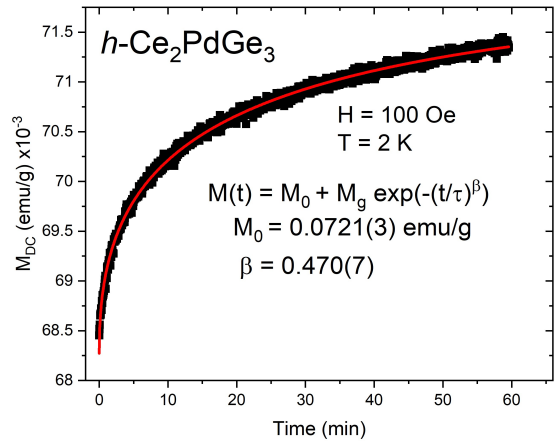

Fig. S5. Time dependent remnant magnetization for $h$-Ce$_2$PdGe$_3$ measured in FC mode at $T = 2$ K. Red line represents a fit to the stretched exponential law: $M = M_0 - M_g\exp(-(t/\tau)^\beta)$.